\documentclass[aps,prd,a4paper,twocolumn,showpacs,nofootinbib]{revtex4}
\usepackage{amssymb,amsmath,latexsym,mathrsfs}
\usepackage{graphicx}
\usepackage{epsfig}
\usepackage{varioref,xr-hyper}
\usepackage{color}

\usepackage[hypertex,draft=false,verbose=false,debug=false
            hyperindex=true,hypertexnames=true]{hyperref}

\newcommand{\hide}[1]{{}}

\newcommand{\EA}{\textsl{et al.~}}

\newcommand{\ep}{\epsilon}

\newcommand{\Om}{\Omega}
\newcommand{\si}{\sigma}
\newcommand{\be}{\begin{equation}}
\newcommand{\ee}{\end{equation}}
\newcommand{\bea}{\begin{eqnarray}}
\newcommand{\eea}{\end{eqnarray}}

\def\lsim{\;\raise 0.4ex\hbox{$<$}\kern -0.8em\lower 0.62 ex\hbox{$\sim$}\;}
\def\gsim{\;\raise 0.4ex\hbox{$>$}\kern -0.7em\lower 0.62 ex\hbox{$\sim$}\;}

\pacs{26.35.+c, 98.80.Cq, 98.80.Ft \hfill CERN-PH-TH/2008-026}

\begin{document}

\title{Is Cosmology Compatible with Blue Gravity Waves ?}
\author{Roberta Camerini}
\affiliation{Physics Department, Universita' di Roma ``La Sapienza,
  Ple Aldo Moro 2, 00185, Rome, Italy}
\author{Ruth Durrer}
\email{ruth.durrer@physics.unige.ch}
\affiliation{D\'epartement de Physique Th\'eorique, Universit\'e de
Gen\`eve, 24 quai Ernest Ansermet, 1211 Gen\`eve 4, Switzerland.}
\author{Alessandro Melchiorri}
\email{alessandro.melchiorri@roma1.infn.it}
\affiliation{Physics Department and INFN, Universita' di Roma ``La
  Sapienza, Ple Aldo Moro 2, 00185, Rome, Italy}
\author{Antonio Riotto}
\email{riotto@cern.ch}
\affiliation{CERN, Theory Division, Gen\`eve 23, CH-1211, Switzerland\\
INFN, Sezione di Padova, via Marzolo 8, 35131 Padova, Italy}
\date{\today}

\begin{abstract}
\noindent
A  primordial gravitational wave background with positive
(blue) spectral index is expected in several non-standard
inflationary cosmologies where the stress-energy tensor violates the null energy
condition. Here we show that a sizable amount of blue gravitational
waves is compatible with current cosmological and astrophysical data.
So far most of the works on parameter estimation from cosmic
microwave background data have assumed a negative
or negligible spectral index. The present limits on cosmological parameters,
especially on the scalar spectral index, widen up considerably when one
allows also for blue tilts of the tensor spectrum.
Since the amplitude of the CMB B-mode polarization is larger in these
models, future data from Planck are likely to provide crucial 
measurements.
\end{abstract}


\maketitle

\section{Introduction}

During the last decade or so, the fluctuations and the polarization of the
cosmic microwave background (CMB) have proven to be the most valuable
observational tool to constrain cosmological models
(see e.g.  ~\cite{wmap3cosm}).
To a big extent this is due to that fact that CMB
physics is sufficiently simple so that it can be calculated to high
accuracy with moderate computational investment. Therefore, accurate
experimental results can be compared with accurate theoretical
predictions.
CMB observations are so valuable, since
they provide a window to the physics at inflation
(see e.g. \cite{alabidi,Peiris:2006ug,Parkinson:2006ku,
Pahud:2006kv,Lewis:2006ma,Seljak:2006bg,Magueijo:2006we,Easther:2006tv,kinney06,
friedman06,finelli06}).

This physics most
probably involves the highest energies ever probed by an experiment,
more the ten orders of magnitude higher than energies achieved in
terrestrial experiments like LHC at CERN. An important prediction of
simple inflationary models is the production of a stochastic
background of gravity waves (\cite{GWs})
with a slightly tilted spectrum,
\be
n_T = -2\ep ~,
\ee
 where $\ep=-\dot{H}/H^2$ denotes a slow roll parameter from
 inflation and $H$ is the Hubble rate during the inflationary stage.
Since in standard inflation $\ep$   is strictly positive~\cite{lr},  in
 the usual  parameter estimation routines, the tensor spectral index
is assumed to be ``red'' ($n_T\le0$, see e.g. \cite{Peiris:2006ug,kinney06})
or negligible~\cite{wmap3cosm}.
However, this does not rule out a priori
the possibility that the spectral index of tensor modes might be
positive, $n_T>0$, i.e. ``blue''.
For instance, in the string gas cosmology set-up \cite{stringgas}, where
scalar metric perturbations are thought to originate from initial string
thermodynamic fluctuations \cite{stringpert},
a spectrum of blue gravity waves (BGW hereafter) is predicted \cite{blue}.
The latter is also a generic prediction of a class of four-dimensional
models characterized by a bouncing phase of the universe. To induce the
bounce, the stress-energy tensor must violate the null energy condition
(NEC). In a spatially flat FRW metric, the NEC corresponds to the inequality
$\dot{H}<0$ and is ultimately responsible for the red tensor spectrum
in standard inflation. In the class of bouncing models \cite{bounce}
the scalar metric perturbations are originally 
of isocurvature nature and they are
subsequently transformed into adiabatic ones. 
The violation of the NEC allows
a BGW spectrum. The same is true in the so-called super-inflation
models \cite{super} where inflation is driven by a component violating 
the NEC. BGW may occur also in scalar-tensor theories and $f(R)$
gravity theories.

While we are aware that all the scenarios
mentioned so far are not  free from criticisms
\cite{lindegas,lindebounce}, due to our ignorance of the dynamics of
the very early universe, in this paper we will assume a
more phenomenological  attitude accepting the possibility that tensor modes
might have a blue spectrum.
This option, together with the rather surprising fact that, as far as
we know, this parameter range has not yet been fully analyzed,
prompted us to investigate whether the presence of such
a BGW spectrum is compatible with current
cosmological and astrophysical observations\footnote{A notable
exception is the paper
\cite{finelli06}. In this work the authors considered a BGW but with an upper limit
$n_T< 0.2$ which, as we will see, is much smaller than
the range of values studied here.}. Moreover, a BGW
spectrum can affect the constraints
on other cosmological parameters which have recently been
derived from present observations of the
cosmic microwave background.
It is not surprising that the limit on the tensor to scalar ratio
$r$ depends on this assumptions, but as we shall see below, also other
quantities like the scalar spectral index, $n_s$, the physical
baryon density $\Om_bh^2=\omega_b$ and the cold dark matter density,
$\Om_ch^2=\omega_c$ do. (Here $\Om_b$ is the
baryon density parameter, $\Om_c$ is the cold dark matter density
parameter and $h$ is the value of the Hubble constant in units of
$100$km/sec/Mpc.)

A careful study of the current astrophysical constraints
on the spectral index of the gravity wave background has been
recently presented in \cite{SB}.
If we assume that the stochastic gravity wave spectrum which affects
CMB fluctuations, at wave lengths of the order of the Hubble scale
from $H_0^{-1}\simeq 3\times 10^{17}h^{-1}$sec to about $10^{-2}H_0^{-1}$,
extends up to wavelengths of about $10^9$sec relevant for the
timing of millisecond pulsars, constraints of the order of $n_T\lsim
0.53$ can be found~\cite{jenet05,jenet06,SB}.
Similar constraints can be obtained
 from the LIGO interferometer (see \cite{abbott06})
while applying the nucleosynthesis
bound on a gravity wave background yields the best constraint,
$n_T\lsim 0.15$. However, these
limits have been obtained by assuming the CMB upper limit on
the scalar/tensor ratio of primordial perturbations
(taken at wavelength $k=0.002 hMpc^{-1}$) $r<0.3$ taken from
the recent WMAP+SDSS analysis of \cite{wmap3cosm} which assumes $n_T=0$.
Since a correlation clearly exists between the tensor
amplitude and the spectral index (e.g., clearly no
constraint on $n_T$ can be derived if $r$ is negligible),
here we provide a further analysis by properly analyzing the full
CMB data and correlations. Moreover, those limits apply only if the
rather bold  extrapolation is made that the
gravity wave spectrum has a constant spectral slope $n_T$
over the range of many orders of magnitude. The
wavelength relevant for LIGO is $\sim 10$sec and for
nucleosynthesis even a fixed spectral index up to the Planck scale,
$\sim 10^{-43}$sec, is
assumed to obtain the above rather stringent limits~\cite{abbott06,SB}.
Since these extrapolations are so bold (we are not aware of any
physical situation where a scaling behavior extends over more than ten
orders of magnitude), we shall also study the case where the limits
derived in Ref.~\cite{SB} do not apply.

Below we also show how the limits on other parameters are affected when we
allow for BGW spectra. We investigate the 3-year
WMAP data~\cite{3years} and we produce forecasts for the
very near future Planck~\cite{planck} satellite experiment.

\section{Analysis Method and Results}

\begin{figure}
\includegraphics[height=8cm,angle=-90]{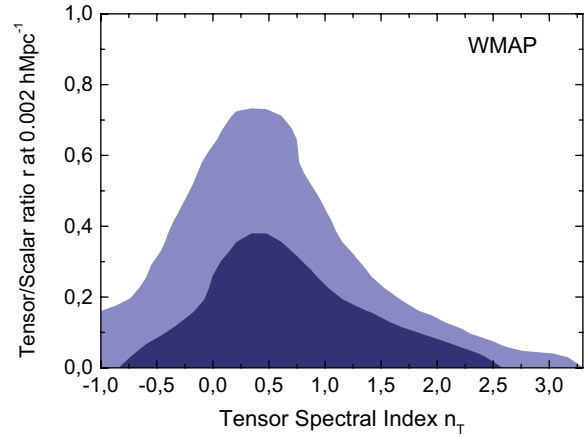}
\caption{\label{f:roby}
$68 \%$ and $95 \%$ likelihood contour plots from WMAP data where
no external prior on the spectral index of the gravity wave
background is assumed. }
\end{figure}

\begin{figure}
\includegraphics[height=6cm]{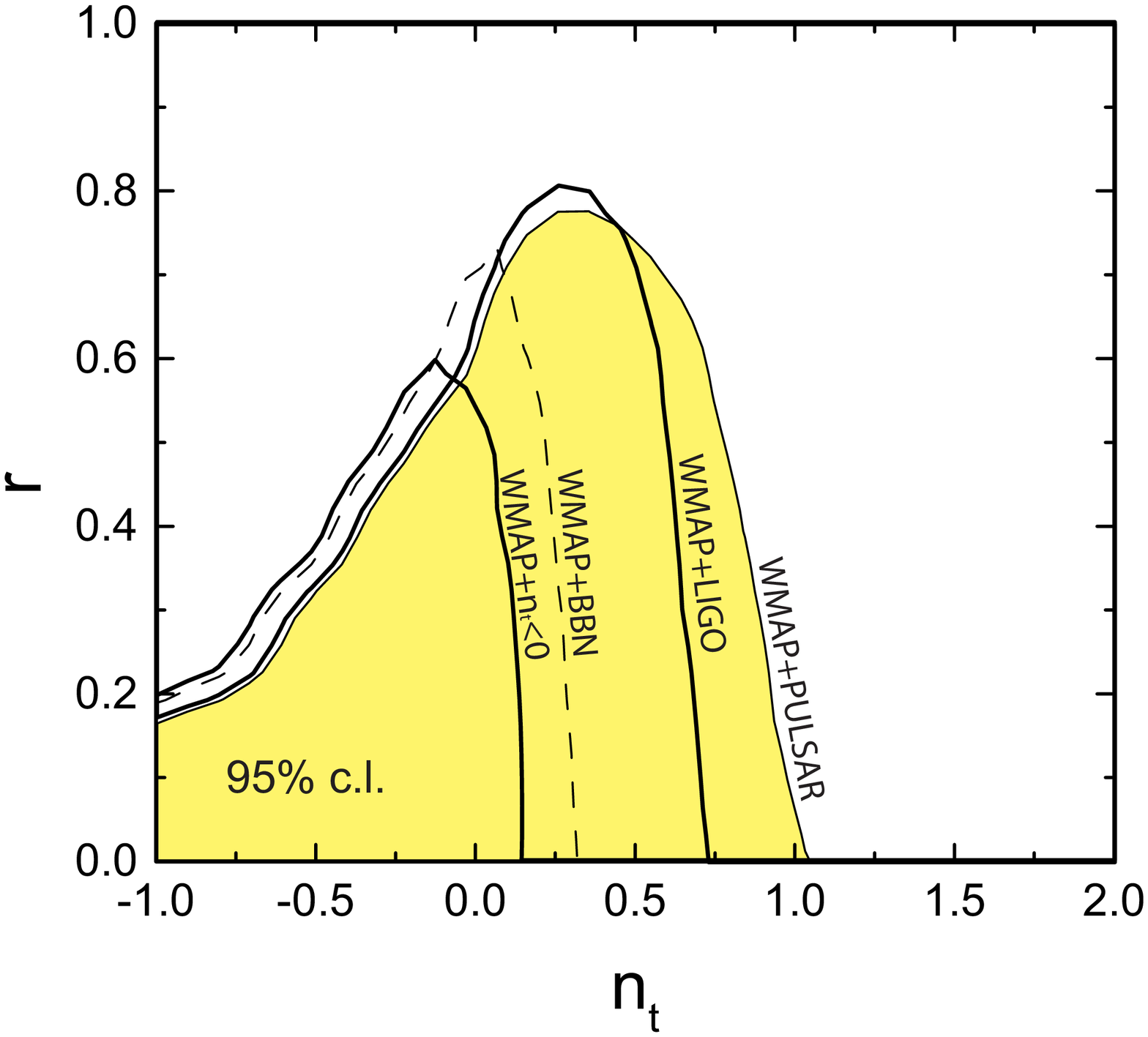}
\caption{\label{f:wmap1}
$95 \%$ likelihood contour plots on the $n_T$-$r$ plane
from WMAP data when external
prior on the spectral index of the gravity wave background are
assumed.}
\end{figure}

\begin{figure}
\includegraphics[height=6cm]{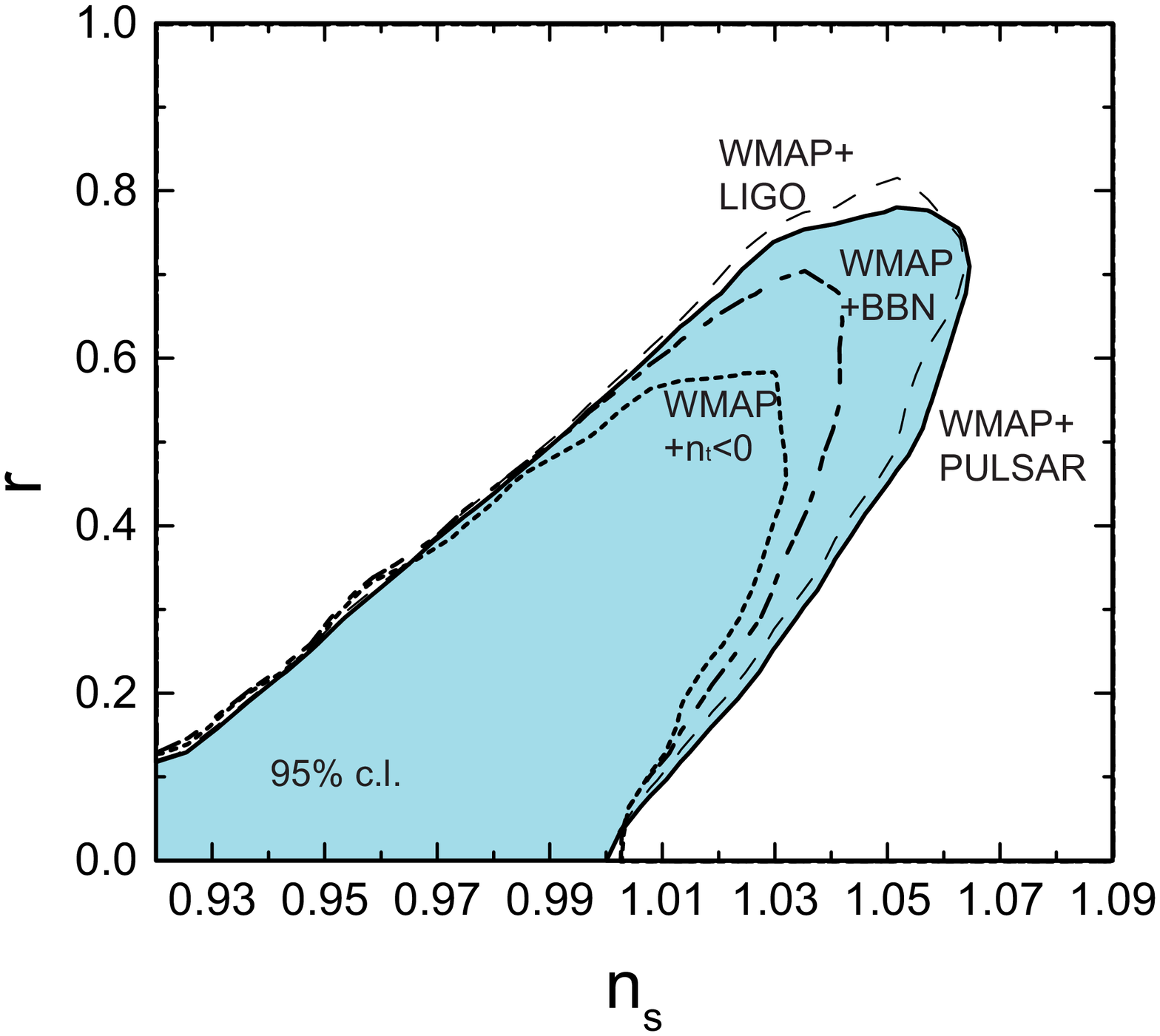}
\caption{\label{f:wmap2}
$95 \%$ likelihood contour plots on the $n_S$-$r$ plane
from WMAP data when external
prior on the spectral index of the gravity wave background are
assumed.}
\end{figure}

\begin{figure}
\includegraphics[height=6cm]{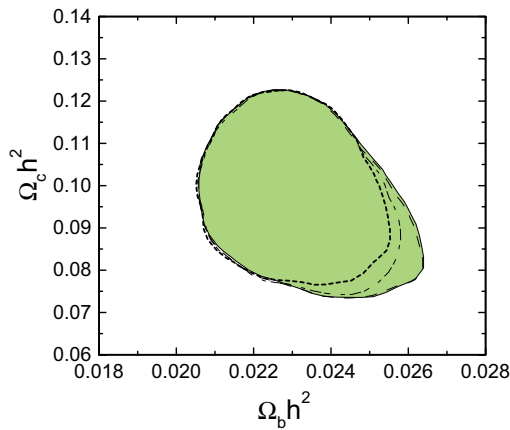}
\caption{\label{f:wmap3}
$95 \%$ likelihood contour plots on the $\omega_b$-$\omega_c$ plane
from WMAP data when external
priors on $n_T$ are
assumed. The solid line corresponds
to the WMAP+PULSAR, the dashed line to WMAP+LIGO, the
long-short dashed to WMAP+BBN and the dotted line to
WMAP+$n_T<0$ case.}
\end{figure}

The method we adopt is based on the publicly available Markov Chain Monte Carlo
package \texttt{cosmomc} \cite{Lewis:2002ah} with a convergence
diagnostics done through the Gelman and Rubin statistics.
We sample the following eight-dimensional set of cosmological
parameters, adopting flat priors on them: the baryon and cold dark
matter densities $\omega_b$ and $\omega_c$,
the ratio of the sound horizon to the angular diameter
distance at decoupling, $\theta_s$, the scalar spectral index
$n_S$, the overall normalization of the spectrum $A$ at $k=0.002$
Mpc$^{-1}$, the optical depth to reionization, $\tau$, the
tensor-to-scalar ratio $r$  at $k=0.002$
Mpc$^{-1}$ and, finally, the tensor spectral index $n_T$.
Furthermore, we consider
purely adiabatic initial conditions and we impose spatial flatness.

We include the three--year WMAP data \cite{wmap3cosm} (temperature
and polarization) with the routine for computing the likelihood
supplied by the WMAP team. Moreover we also include in a separate
analysis the new, high quality, fine-scale measurements from
the ACBAR experiment (\cite{acbar07}).

We have studied three different possible priors on
the combination between $n_T$ and $r$ as derived in
Ref.~\cite{SB} from pulsar timing, LIGO and
nucleosynthesis with the Planck scale as the upper cutoff:

\bea
n_T &\le & 0.0477\log\left(\frac{2.09\times 10^4}{r}\right) \quad
\mbox{PULSAR,} \label{pul}\\
n_T &\le & 0.0223\log\left(\frac{1.49\times 10^{10}h^2}{r}\right) \quad
\mbox{LIGO,} \\
n_T &\le & 0.00714\log\left(\frac{3.4\times10^9h^2}{r}\right) \quad
\mbox{ BBN.}  \label{nuc}
\eea

It is important to notice the dependence of the
LIGO and BBN constraints on $h$.

In addition to this astrophysical constraints we have also simply considered a prior
$-1<n_T<20$ without taking the above limits into account, since it may
very well be that the gravity wave spectrum is modified on some scale
below those needed for the limits of Eqs.~(\ref{pul}) to (\ref{nuc}).
The results are also compared with a $2-\sigma$ upper limit of $n_T<0$

A first analysis is made in the theoretical framework described
above with no upper limits on  $n_T$. The likelihood
contour plot in the $n_T$-$r$ plane is reported in Figure~\ref{f:roby}.
When only the WMAP data is considered, no
upper limit on $n_T$ alone can be derived. The reason is simple:
gravity waves with BGW spectra can always be accommodated with
 the WMAP data by lowering the tensor/scalar ratio $r$.
However it is possible to derive the following
 $95 \%$ upper limits on $n_T$ in function of $r$:
$n_T<2.1$ for $r>0.1$ and $n_T<3.2$ for $r>0.01$.
In the range $0.05 < r < 0.6$ and $0.75 < n_T < 3$ we derive
the following upper limit fit at $95 \%$ c.l. from WMAP only:

\begin{equation}
r+0.68n_T-0.12n_T^2 \le 1
\end{equation}

\noindent while including the ACBAR data we obtain:

\begin{equation}
r+0.41n_T-0.06n_T^2 \le 0.7
\end{equation}

\noindent again at $95 \%$ c.l..
Assuming $r>0.05$ and no prior on $n_T$, we found from
WMAP+ACBAR $-1.51< n_T < 2.62$ at $95 \% c.l.$

These limits, while new and on
very different scales, are however rather weak if compared
with the upper limits described in Eqs.~(\ref{pul}) to~(\ref{nuc}) above.
It is however interesting to notice that even when these
stronger limits are considered, BGW spectra are
still compatible with CMB observations. Since this
contribution has not been usually considered in
CMB parameter analysis, it is interesting to study the possible
effect of such a BGW background.

In Figure~\ref{f:wmap1},~\ref{f:wmap2} and ~\ref{f:wmap3}
 we show how cosmological parameters as
inferred from the WMAP 3-year data are
affected by a BGW spectrum.

In Figure~\ref{f:wmap1} we see how current external
priors on $n_T$ affect the WMAP bounds on $r$.
As we can see, current bounds on $r$ obtained for $n_T<0$ can
be relaxed by as much as a factor $1.4$
when BGW spectra in agreement with LIGO and PULSAR data are
assumed. In particular we found the following $95 \%$ c.l.
upper limits on $r$: $r<0.64$, $r<0.65$, $r<0.55$ and $r<0.47$
when the WMAP data is combined with the PULSAR, LIGO, BBN and
$n_T<0$ priors respectively. We found that the constraints
with no external prior such that $n_T<1$ are very similar to
the PULSAR and LIGO cases.

In Figure~\ref{f:wmap2} we see how the parameter
ranges for the pair $(n_S,r)$ are affected by a blue tilt in the
tensor spectrum. First, it is interesting to note that a larger tensor
contribution is allowed and the scalar spectral index can then be
bluer. This comes from the well known partial degeneracy between
rising the scalar spectral index which reduces the Sachs Wolfe plateau
if the height of the first acoustic peak is fixed, and at the same
time rising the tensor contribution which enhances the Sachs Wolfe plateau.
This degeneracy is strengthened if $n_T>0$ since then tensor
fluctuations in the CMB extend to somewhat smaller scales.
Disregarding the very speculative nucleosynthesis prior, the $2\si$
limit on the scalar spectral index is enhanced from $n_s=0.981\pm0.026$ to
 $n_s=0.994\pm0.032$ when allowing for BGW spectra consistent
with pulsar timing. BGW's affect also the constraints on
the optical depth parameter, which moves from $\tau=0.091\pm0.030$ to
 $\tau=0.097\pm0.032$ for the WMAP+PULSAR case.

As one sees in Figure~\ref{f:wmap3}, enhancing the scalar
spectral index is slightly correlated with enhancing $\omega_b$ which
leads to stronger Silk damping. The higher baryon density is
compensated by a lower cold dark matter
density. This not in a way so that $\omega_m=\omega_c+\omega_b$ would
remain constant, but so that the total matter density decreases which
also reduces the acoustic peaks, to compensate for the blue
spectral index. The $2\si$ limits on $\omega_c$ and $\omega_b$ change from
$\omega_b=0.0228\pm0.0009$ and $\omega_c=0.100\pm0.009$
 to $\omega_b=0.0232\pm0.0011$ and $\omega_c=0.098\pm0.01$
 respectively when allowing for blue tensor spectra consistent
with pulsar timing. The effect of BGW's is therefore
smaller on those parameters but still measurable.

\begin{figure}
\includegraphics[height=9cm]{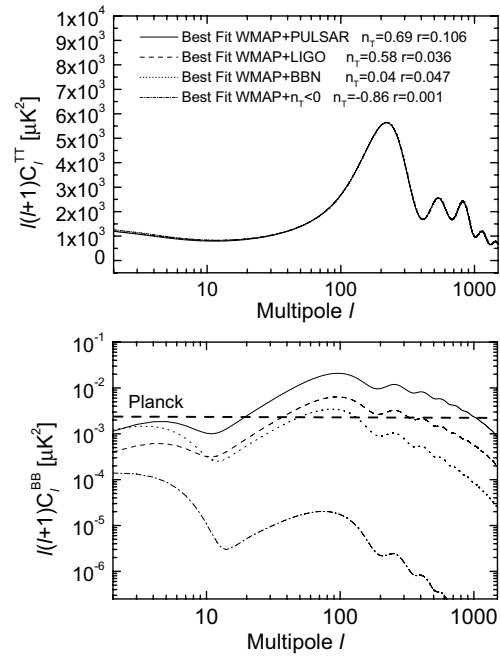}
\caption{\label{f:cielle}
Top Panel: CMB Temperature anisotropy angular power spectra
for the best fit models to current WMAP data for different
choices of external priors on $n_T$.
Bottom Panel: Contribution to the $B$ mode polarization.}
\end{figure}

While current CMB constraints are clearly affected by
the inclusion of a blue gravity wave component it is
interesting to investigate the impact on future measurements
as those expected from the Planck satellite.
The first important observation is that a blue tensor
component dramatically enhances the probability for the detection of a
gravity wave background with
 PLANCK. In Figure~\ref{f:cielle} we plot the best fit
angular anisotropy power spectra to the WMAP data for
the different external priors on $n_T$. As we can see,
all the models give nearly identical power spectra, i.e.
it is impossible to discriminate between those model with current
CMB observations.

CMB polarization is, however, a very promising tool for
detecting a BGW background. The statistical properties of CMB
linear polarization are fully characterized by two
sets of spin-2 multipole moments with opposite parities
\cite{szeb}. As well known in the literature
(see e.g. \cite{marctwo,pagano}) the
magnetic-type modes (B or curl modes) are produced by
tensor metric perturbations and
not scalar perturbations.
A detection of B-mode polarization would thus
provide good evidence for a primordial background
of gravity waves.

As we can see from the bottom panel of
Figure~\ref{f:cielle} the expected amplitude of the polarization
$B$ mode differs dramatically between the models. In particular, it is
important to notice that a blue spectral index could enhance the
$B$ mode power at $\ell \sim 100$ by several orders of magnitude
if compared with models with $n_T\le0$. Also in the plot we also
report the expected sensitivity from the Planck experiment
(dashed horizontal line). Most of the BGW spectra models
are above the sensitivity of PLANCK.

We have  simulated future cosmological data with a noise
spectrum based on the expected Planck configuration. In particular
we assume an experimental sensitivity $w_p^{-1/2}=81 \mu K$,
beam size $\theta_{FWHM}=7.1'$ and sky coverage $f_{sky}=0.8$
We also assume a spatially flat model with parameters $h=0.7$,
$\omega_c=0.120$, $\omega_b =0.022$, $n_S=1$ and $r=0$.
We find that the constraints on tensor modes expected
for this model will be equally affected.
The $2\si$ limit on $r$ is indeed relaxed by $\sim 30 \%$
from $r < 0.025$ to $r < 0.035$ when allowing for blue tensor spectra
consistent with pulsar timing.

\section{Conclusions}

In this paper we have analyzed to which extent a primordial
background of gravitational waves with positive spectral index is
compatible with current observations.
We have found that a considerable part of the parameter space of
cosmological models
is in agreement with current WMAP data and with upper limits
coming from the LIGO experiment and from pulsar timing.
Several non standard models can produce such a background and
there is therefore no strong theoretical reason to exclude these models in
current parameter estimation analysis. With this in mind, we have shown
that, if the $n_T<0$ assumption is relaxed,
also parameters which are not directly related to the
tensor component are affected. Especially, the scalar spectral index
can be significantly bluer than in models which allow only for a red
tilt of the tensor spectrum. Moreover, we have found a smaller but 
significant effect on the baryon and cold dark matter density constraints.
This example shows once more that when assuming any
limits on cosmological parameters it is very
important to be aware of the model assumptions which went into their
derivation.

Future Planck estimates can also be affected by BGW spectra,
especially when constraints on $r$ are considered.
However they will be nearly entirely free of the additional
degeneracies which are still seen in the WMAP data.
As a final remark, we like to point out that blue spectra are
able to produce significantly larger contribution to the $B$-mode
polarization spectrum. An excess of $B$-mode polarization at $\ell
\sim 100$ would therefore provide convincing evidence for
the BGW models investigated here.

\vspace{0.6cm}

\noindent {\bf Acknowledgment}\\
We thank Martin Kunz for discussions. RD is supported by the Swiss
National Science Foundation. RC thanks Geneva University for
hospitality. AM thanks the CERN Theory Division for financial support
during the early stages of this work.
This research was supported in part by the European Community's Research
Training Networks under contracts MRTN-CT-2004-503369 and MRTN-CT-2006-035505.
This research has been supported by ASI contract I/016/07/0 "COFIS".

\end{document}